\documentclass[a4paper,11pt,leqno,twoside,intlimits,namelimits,sumlimits]{myarticle}

\usepackage{amsmath,amssymb,fancyhdr,latexsym,theorem}

\allowdisplaybreaks

\newcommand{\fin}{f^{\text{in}}}
\newcommand{\phinin}{\phi_0^{\text{in}}}
\newcommand{\phiein}{\phi_1^{\text{in}}}
\newcommand{\Deltain}{\Delta^{\text{in}}}
\newcommand{\ret}{\big|_{\text{ret}}}

\newcommand{\Ret}{\Big|_{\text{ret}}}
\newcommand{\SP}{\!\cdot\!}
\newcommand{\negsp}{\!\!}

\DeclareMathOperator{\supp}{supp}
\DeclareMathOperator{\meas}{meas}
\DeclareMathOperator{\divergence}{div}

\newcommand{\R}{\mathbb{R}}

\theoremstyle{plain}
\theoremheaderfont{\upshape\bfseries}
{\theorembodyfont{\slshape}
  \newtheorem{theorem}{Theorem}
  \newtheorem{definition}{Definition}
  \newtheorem{lemma}{Lemma}
  \newtheorem{proposition}{Proposition}
  
}
{\theorembodyfont{\upshape}
{

}

\newenvironment{proof}
{
{\slshape Proof:{ }}
}
{
\hfill$\Box$\par\vspace*{0.2cm}
}

\pagestyle{fancy}
\fancyfoot{}
\fancyhead{}

\fancyhead[EL]{{\small\scshape\thepage}}
\fancyhead[EC]{{\small\scshape Stefan Friedrich}}
\fancyhead[OC]{{\small\scshape Global Small Solutions of the Vlasov-Nordstr\"om System}}
\fancyhead[OR]{{\small\scshape\thepage}}

\begin{document}
\title{\vspace*{-1cm}{\bfseries\sffamily Global Small Solutions of the Vlasov-Nordstr\"om System}}
\author{{\scshape Stefan Friedrich}\\ {\normalsize Department of Mathematics, University of Bayreuth}\\ {\normalsize 95440 Bayreuth, Germany}}
\date{{\normalsize\today}}
\maketitle

\begin{abstract}
The Vlasov-Nordstr\"om system is a relativistic model for the description of a self-gravitating collisionless gas. In this paper we show, using a bootstrap argument, that classical small solutions of the Vlasov-Nordstr\"om system exist globally in time.
\end{abstract}

\section{Introduction}
The Vlasov-Nordstr\"om system serves for the description of galaxies and globular clusters under the influence of gravitation. If one neglects collisions between particles and uses the Nordstr\"om scalar theory of gravitation coupled to the Vlasov equation, one arrives at the so called Vlasov-Nordstr\"om system, which is more complicated than the non-relativistic Vlasov-Poisson system based on Newtonian theory of gravitation and much easier than the Vlasov-Einstein system of general relativity.

Let us denote by $f(t,x,p)$ the density function of the particles on phase space, where $t\ge 0$ denotes time and $x,p\in\R^3$ position and momentum, respectively. If we denote the scalar field of gravitation by $\phi(t,x)$, the system reads
\begin{eqnarray}
\partial_t^2\phi-\bigtriangleup_x\phi=-\int\frac{f(t,x,p)}{\sqrt{1+p^2}}\,dp=:-\mu(t,x),\label{waveequation}\\
Sf-\left[(S\phi)p+(1+p^2)^{-1/2}\partial_x\phi\right]\SP\partial_pf=4fS\phi,\label{vlasovequation}
\end{eqnarray}
where $S:=\partial_t+\hat{p}\SP\partial_x$ is the free-transport operator and $\hat{p}:=p/\sqrt{1+p^2}$ is the relativistic velocity of a particle with momentum $p$. Integrals without domain of integration extend always over $\R^3$. In equations (\ref{waveequation}) and (\ref{vlasovequation}) we have chosen the units such that the mass of each particle, the gravitational constant and the speed of light are all equal to unity.

For a more detailed discussion of the physical background of the involved quantities $\phi$ and $f$ as well as the differential geometric interpretation we refer to \cite{andreassoncalogerorein,calogero03,calogerorein03,calogerorein2}.

We supply the system with the initial conditions
\begin{equation}\label{initialconditions}
f(0,x,p)=\fin(x,p),\,\phi(0,x)=\phinin(x),\,\partial_t\phi(0,x)=\phiein(x),\quad x,p\in\R^3,
\end{equation}
and we assume that they have the regularity
\begin{equation}\label{regularityofinitialconditions}
0\le\fin\in C_c^1(\R^6),\quad\phinin\in C_c^3(\R^3),\quad\phiein\in C_c^2(\R^3).
\end{equation}
The subscript $c$ indicates compact support of the functions under consideration.

For the rest of the paper let us fix some $R>0$. As space of initial conditions we use
\begin{align*}
\mathcal{I}:=\mathcal{I}_R:=\{(\fin,\phinin,\phiein)\,&|\,\fin,\phinin,\phiein\text{{ }as in (\ref{regularityofinitialconditions}),{ }}\Deltain\le 1,\\
&\,\supp\fin\subset B_R(0)\subset\R^6,\\
&\,\supp\phinin,\supp\phiein\subset B_R(0)\subset\R^3\},
\end{align*}
equipped with the norm
\begin{displaymath}
\Deltain:=||(\fin,\phinin,\phiein)||:=||\fin||_{1,\infty}+||\phinin||_{3,\infty}+||\phiein||_{2,\infty}.
\end{displaymath}
Here we have denoted by $B_R(0)$ the ball with radius $R$ centered at the origin and by $||\cdot||_{k,\infty}$ the sum of the $L^\infty$-norms of the derivatives up to order $k$.

The theorem on global existence of small solutions can now be stated:
\begin{theorem}\label{maintheorem}
There exists a $\delta=\delta(R)>0$ such that for all $(\fin,\phinin,\phiein)\in\mathcal{I}$ with $\Deltain<\delta$ the unique, classical solution of (\ref{waveequation}), (\ref{vlasovequation}) and (\ref{initialconditions}) exists on $[0,\infty[$, and for all $t\ge 0$ and $x\in\R^3$ with $|x|\le R+t$ we have
\begin{align*}
|\partial_t\phi(t,x)|,|\partial_x\phi(t,x)|&\le C(1+R+t+|x|)^{-1}(1+R+t-|x|)^{-1},\\
&\hspace*{-3.43cm}|\partial_t^2\phi(t,x)|,|\partial_t\partial_x\phi(t,x)|,|\partial_x^2\phi(t,x)|\\
&\le C(1+R+t+|x|)^{-1}(1+R+t-|x|)^{-7/4},\\
||\mu(t)||_\infty&\le C(1+t)^{-3}.
\end{align*}
\end{theorem}

This theorem will be proved by a standard bootstrap argument in connection with a continuation criterion, which has been proved in \cite{calogerorein03,calogerorein2}.

By the continuation criterion, initial data as specified in (\ref{regularityofinitialconditions}) launch a unique classical solution on a maximal time interval $[0,T_{max}[$. If furthermore
\begin{displaymath}
\sup\{|p|\,|\,(x,p)\in\supp f(t), 0\le t<T_{max}\}<\infty,
\end{displaymath}
then $T_{max}=\infty$.

In the following, by solutions we always mean classical solutions in the sense of \cite{calogerorein03,calogerorein2}.

Let us introduce some notations. By $C$ we denote a constant, which may change from line to line, but which does not depend on the initial data. By $(X(s,t,x,p),P(s,t,x,p))$ we denote, as usual, the solution of the characteristic equations
\begin{displaymath}
\frac{d}{ds}x=\hat{p},\qquad\frac{d}{ds}p=-(S\phi)p-(1+p^2)^{-1/2}\partial_x\phi,
\end{displaymath}
corresponding to (\ref{vlasovequation}) with initial conditions $x(t)=x$, $p(t)=p$.
For the sum of the absolut values of the first or second derivatives of $\phi$ we write for short
\begin{align*}
K(t,x)&:=|\partial_t\phi(t,x)|+|\partial_x\phi(t,x)|,\\
L(t,x)&:=|\partial_t^2\phi(t,x)|+|\partial_t\partial_x\phi(t,x)|+|\partial_x^2\phi(t,x)|.
\end{align*}
It is well known that $f$ admits a representation as
\begin{equation}\label{frep}
f(t,x,p):=\fin(X(0,t,x,p),P(0,t,x,p))e^{4\phi(t,x)-4\phinin(X(0,t,x,p))}.
\end{equation}
The compact support property of $f$ in connection with the fact that $|\hat{p}|<1$ yields
\begin{equation}\label{fzero}
f(t,x,\cdot)=0\qquad\text{for{ }}|x|>R+t.
\end{equation}
From Proposition 1 in \cite{calogerorein2} and elementary estimates of the solution $\phi_0$ of the homogeneous wave equation (cf.~the proof of Lemma \ref{representationfirstt}) we know
\begin{equation}\label{fest}
||f(t)||_\infty\le||\fin||_\infty e^{C(||\phinin||_{1,\infty}+||\phiein||_\infty)}\le C\Deltain\le C,
\end{equation}
where the last two inequalities are only valid for $(\fin,\phinin,\phiein)\in\mathcal{I}$.

\section{Properties of Free-Streaming Solutions}\label{freestreamingsection}

The small solutions we have in mind satisfy a special kind of decay condition, so we start with

\begin{definition}
Let $(f,\phi)$ be a solution of (\ref{waveequation}), (\ref{vlasovequation}) on some time interval $[0,T[$, $T>0$. We say that $(f,\phi)$ satisfies a free-streaming condition (FSC) with respect to $\eta>0$ on $[0,T[$, if for all $t\in[0,T[$ and all $x\in\R^3$ with $|x|\le R+t$ we have
\begin{align*}
K(t,x)&\le\eta(1+R+t+|x|)^{-\beta}(1+R+t-|x|)^{-\beta},\\
L(t,x)&\le\eta(1+R+t+|x|)^{-\beta}(1+R+t-|x|)^{-\beta-1},
\end{align*}
for some $\beta\in]\frac{1}{2},\frac{3}{4}[$.
\end{definition}

We derive now some properties of free-streaming solutions with sufficiently small $\eta$.

\begin{lemma}\label{supportestimate}
There exists an $0<\eta_1\le 1$ such that the following holds: If $(f,\phi)$ is a solution of (\ref{waveequation}), (\ref{vlasovequation}) and (\ref{initialconditions}) on some time interval $[0,T[$, $T>0$, with $(\fin,\phinin,\phiein)\in\mathcal{I}$ that satisfies (FSC) with respect to $\eta<\eta_1$, then we have
\begin{displaymath}
\supp f(t)\subset\{(x,p)\in\R^3\times\R^3\,|\,|x|\le R+\hat{C}t, |p|\le 2R\},\quad t\in[0,T[,
\end{displaymath}
where $\hat{C}:=(2R)(1+(2R)^2)^{-1/2}$.
\end{lemma}
\begin{proof}
At first, let $\eta>0$ be arbitrary and $(f,\phi)$ be a solution of (\ref{waveequation}), (\ref{vlasovequation}) and (\ref{initialconditions}) satisfying the assumptions of the lemma. Obviously,
\begin{displaymath}
\supp f(t)=\{(X(t,0,x,p),P(t,0,x,p))\,|\,(x,p)\in\supp f(0)\}.
\end{displaymath}
Consider now an $(x,p)\in\supp f(0)$ and define
\begin{displaymath}
\tilde{t}:=\sup\{t\in[0,T[\,|\,|P(s,0,x,p)|\le 2R, s\in[0,t]\}.
\end{displaymath}
Then we have $\tilde{t}>0$ and we get from the characteristic equations
\begin{displaymath}
|X(t,0,x,p)|\le R+\int_0^t\frac{|P(s,0,x,p)|}{\sqrt{1+P^2(s,0,x,p)}}dx\le R+\hat{C}t,\quad t\in[0,\tilde{t}[,
\end{displaymath}
$\hat{C}$ as proposed in the lemma. From the characteristic equation for $p$ we obtain
\begin{equation}\label{Pest}
|P(t,0,x,p)|\le R+C(1+2R)\int_0^t|K(s,X(s))|\,ds,\quad X(s):=X(s,0,x,p),
\end{equation}
again for $t\in[0,\tilde{t}[$. The free-streaming condition yields
\begin{align*}
|K(s,X(s))|&\le\eta(1+s+|X(s)|)^{-\beta}(1+R+s-|X(s)|)^{-\beta}\\
&\le\eta(1+s)^{-\beta}(1+R+s-R-\hat{C}s)^{-\beta}\le C\eta(1+s)^{-2\beta}.
\end{align*}
Substituting this in (\ref{Pest}), we get by integration
\begin{displaymath}
|P(t,0,x,p)|\le R+C(1+2R)\eta,\quad t\in[0,\tilde{t}[.
\end{displaymath}
Now let $\eta_1:=R(2C(1+2R))^{-1}$. Then, $|P(t,0,x,p)|\le\frac{3}{2}R$ for $t\in[0,\tilde{t}[$ and by the definition of $\tilde{t}$ the claim follows.
\end{proof}

\begin{lemma}\label{characteristicestimates}
There exists an $\eta_2>0$, $\eta_2<\eta_1$, such that the following holds: If $(f,\phi)$ is a solution of (\ref{waveequation}), (\ref{vlasovequation}) and (\ref{initialconditions}) on some time interval $[0,T[$, $T>0$, with $(\fin,\phinin,\phiein)\in\mathcal{I}$ that satisfies (FSC) with respect to $\eta<\eta_2$, then we have
\begin{displaymath}
|X(0,t,x,p_1)-X(0,t,x,p_2)|\ge C|p_1-p_2|t,\quad C=C(R)
\end{displaymath}
for $(x,p_1),(x,p_2)\in\supp f(t)$ and $t\in[0,T[$.
\end{lemma}
\begin{proof}
Let $(f,\phi)$ be a solution of (\ref{waveequation}), (\ref{vlasovequation}) and (\ref{initialconditions}) satisfying the assumptions of the lemma. Let for $(x,p_1)$, $(x,p_2)\in\supp f(t)$, $t\in[0,T[$
\begin{displaymath}
x_i(s):=X(s,t,x,p_i),\quad \hat{p}_i(s):=\frac{P(s,t,x,p_i)}{\sqrt{1+P^2(s,t,x,p_i)}},\quad 0\le s\le t, i=1,2.
\end{displaymath}
After a short calculation, we obtain from the characteristic system
\begin{displaymath}
\dot{x}_i(s)=\hat{p}_i(s),\quad\ddot{x}_i(s)=J(s,x_i(s),\hat{p}_i(s))
\end{displaymath}
with
\begin{displaymath}
J(s,x,\hat{p})=-(1-\hat{p}^2)\left[\partial_t\phi(s,x)\hat{p}+\partial_x\phi(s,x)\right].
\end{displaymath}
Now we define for $0\le s\le t$
\begin{displaymath}
y(s):=x_1(s)-x_2(s)+(t-s)(\hat{p}_1-\hat{p}_2)
\end{displaymath}
and obtain $y(t)=0$, $\dot{y}(t)=0$, as well as
\begin{alignat*}{2}
|\ddot{y}(s)|&=|&&J(s,x_1(s),\hat{p}_1(s))-J(s,x_2(s),\hat{p}_2(s))|\\
&=\Bigg|&&\int_0^1\frac{d}{d\tau}J(s,\tau x_1(s)+(1-\tau)x_2(s),\hat{p}_1(s))\,d\tau\\
&&&+\int_0^1\frac{d}{d\tau}J(s,x_2(s),\tau\hat{p}_1(s)+(1-\tau)\hat{p}_2(s))\,d\tau\Bigg|\\
&=\Bigg|&&\int_0^1[\partial_xJ(s,\tau x_1(s)+(1-\tau)x_2(s),\hat{p}_1(s))(x_1(s)-x_2(s))\\
&&&+\partial_{\hat{p}}J(s,x_2(s),\tau\hat{p}_1(s)+(1-\tau)\hat{p}_2(s))(\hat{p}_1(s)-\hat{p}_2(s))]\,d\tau\Bigg|\\
&\le&&S_1(|y(s)|+|t-s||\hat{p}_1-\hat{p}_2|)+S_2(|\dot{y}(s)|+|\hat{p}_1-\hat{p}_2|)
\end{alignat*}
with
\begin{align*}
S_1&=\sup_{0\le\tau\le 1}|\partial_xJ(s,\tau x_1(s)+(1-\tau)x_2(s),\hat{p}_1(s))|,\\
S_2&=\sup_{0\le\tau\le 1}|\partial_{\hat{p}}J(s,x_2(s),\tau\hat{p}_1(s)+(1-\tau)\hat{p}_2(s))|.
\end{align*}
Now we have
\begin{align}
|\partial_{\hat{p}_k}J(s,x,\hat{p})|&\le(1+3|\hat{p}|^2)|\partial_t\phi(s,x)|+2|\hat{p}||\partial_x\phi(s,x)|\nonumber \\
&\le C\,K(s,x),\label{Jpk}\\
|\partial_{x_k}J(s,x,\hat{p})|&\le|\hat{p}^2-1||\hat{p}||\partial_{x_k}\partial_t\phi(s,x)|+|\hat{p}^2-1||\partial_{x_k}\partial_x\phi(s,x)|\nonumber \\
&\le C\,L(s,x).\label{Jxk}
\end{align}
Let $\eta\in]0,\eta_1[$ ($\eta_1$ from Lemma \ref{supportestimate}) be arbitrary. Because of $(x_i(s),\hat{p}_i(s))\in\supp f(s)$ we have from Lemma \ref{supportestimate}, $|x_i(s)|\le R+\hat{C}s$ and $|\hat{p}_i(s)|\le 2R$, in particular
\begin{displaymath}
|\tau x_1(s)+(1-\tau)x_2(s)|\le R+\hat{C}s,\quad |\tau\hat{p}_1(s)+(1-\tau)\hat{p}_2(s)|\le 2R,
\end{displaymath}
and the free-streaming condition finally yields
\begin{align*}
S_1&\le C\sup_{0\le\tau\le 1}L(s,\tau x_1(s)+(1-\tau)x_2(s))\\
&\le C\eta(1+s)^{-\beta}(1+R+s-R-\hat{C}s)^{-\beta-1}\le C\eta(1+s)^{-2\beta-1}
\end{align*}
and
\begin{align*}
S_2&\le C\,K(s,x_2(s))\le C\eta(1+s+|x_2(s)|)^{-\beta}(1+R+s-|x_2(s)|)^{-\beta}\\
&\le C\eta(1+s)^{-\beta}(1+R+s-R-\hat{C}s)^{-\beta}\le C\eta(1+s)^{-2\beta}.
\end{align*}
Hence we have deduced for $y(\cdot)$ the following differential inequality
\begin{alignat*}{2}
|\ddot{y}(s)|&\le &&C\eta(1+s)^{-2\beta-1}|x_1(s)-x_2(s)|+C\eta(1+s)^{-2\beta}|\hat{p}_1(s)-\hat{p}_2(s)|\\
&\le &&C\eta(1+s)^{-2\beta-1}(|y(s)|+|t-s||\hat{p}_1-\hat{p}_2|)\\
&&&+C\eta(1+s)^{-2\beta}(|\dot{y}(s)|+|\hat{p}_1-\hat{p}_2|),\quad 0\le s\le t,
\end{alignat*}
which, together with Lemma 5.4 in \cite{reindiss} implies
\begin{displaymath}
|y(s)|\le|\hat{p}_1-\hat{p}_2|\eta Ie^{\eta I}(t-s),\quad 0\le s\le t.
\end{displaymath}
Here $I=I(\beta)$ is a constant. Choose now an $\eta_2\in]0,\eta_1[$ with $\eta_2Ie^{\eta_2I}<\frac{1}{2}$. Then we have
\begin{displaymath}
|x_1(s)-x_2(s)+(t-s)(\hat{p}_1-\hat{p}_2)|=|y(s)|<\frac{1}{2}|t-s||\hat{p}_1-\hat{p}_2|
\end{displaymath}
respectively
\begin{equation}\label{fastXXfertig}
\frac{1}{2}|t-s||\hat{p}_1-\hat{p}_2|<|X(s,t,x,p_1)-X(s,t,x,p_2)|.
\end{equation}
We calculate
\begin{displaymath}
\partial_{\hat{p}_k}p=\frac{(1-\hat{p}^2)e_k+\hat{p}_k\hat{p}}{(1+\hat{p}^2)^{3/2}},\text{{ }because{ }}p=p(\hat{p})=\frac{\hat{p}}{\sqrt{1-\hat{p}^2}},
\end{displaymath}
and therewith
\begin{align*}
|p_1-p_2|&=\left|\int_0^1\frac{d}{d\tau}p(\tau\hat{p}_1+(1-\tau)\hat{p}_2)\,d\tau\right|\\
&\le\sup_{0\le\tau\le 1}|\partial_{\hat{p}}p(\tau\hat{p}_1+(1-\tau)\hat{p}_2)||\hat{p}_1-\hat{p}_2|\le C|\hat{p}_1-\hat{p}_2|,
\end{align*}
where we used again Lemma \ref{supportestimate} at the last inequality. Substituting this in (\ref{fastXXfertig}) proves the claim.
\end{proof}

\begin{lemma}\label{partialxfestimate}
There exists a  $C>0$, such that the following holds: If $(f,\phi)$ is a solution of (\ref{waveequation}), (\ref{vlasovequation}) and (\ref{initialconditions}) on some time interval $[0,T[$, $T>0$, with $(\fin,\phinin,\phiein)\in\mathcal{I}$ that satisfies (FSC) with respect to $\eta<\eta_2$, then we have
\begin{displaymath}
||\partial_xf(t)||_\infty\le C\Deltain,\quad t\in[0,T[.
\end{displaymath}
\end{lemma}
\begin{proof}
We differentiate (\ref{frep}) with respect to $x$ and estimate, using (\ref{fest}) as follows:
\begin{displaymath}
|\partial_xf(t,x,p)|\le C\Deltain(|\partial_xX(0,t,x,p)|+|\partial_xP(0,t,x,p)|+|\partial_x\phi(t,x)|).
\end{displaymath}
From Lemma \ref{characteristicestimates} we know already that $\ddot{X}(s)=J(s,X(s),\hat{P}(s))$. Hence,
\begin{displaymath}
\partial_{x_k}\ddot{X}(s)=\partial_xJ(s,X(s),\hat{P}(s))\partial_{x_k}X(s)+\partial_{\hat{p}}J(s,X(s),\hat{P}(s))\partial_{x_k}\dot{X}(s).
\end{displaymath}
Now let $x(s):=\partial_xX(s)-I$, $0\le s\le t<T$. Then we have $x(t)=\dot{x}(t)=0$ and $\ddot{x}(s)=\partial_xX(s)$. Using the same arguments as in Lemma \ref{characteristicestimates} and with the help of Lemma \ref{supportestimate} we conclude after a short calculation
\begin{displaymath}
|\ddot{x}(s)|\le C\eta(1+s)^{-2\beta-1}(|x(s)|+1)+C\eta(1+s)^{-2\beta}|\dot{x}(s)|.
\end{displaymath}
Again, we can apply Lemma 5.4 in \cite{reindiss} to obtain
\begin{displaymath}
|x(s)|\le C,\quad\text{thus}\quad|\partial_xX(s,t,x,p)|\le C+1.
\end{displaymath}
To estimate $\partial_xP(s)$ we calculate $\partial_{x_k}\dot{P}$ from the characteristic system and obtain with (\ref{Jpk}), (\ref{Jxk}) and the free-streaming condition after a short calculation
\begin{align*}
|\partial_{x_k}\dot{P}(s)|&\le C\,L(s,X(s))+C\,K(s,X(s))|\partial_{x_k}P(s)|\\
&\le C\eta(1+s)^{-2\beta-1}+C\eta(1+s)^{-2\beta}|\partial_{x_k}P(s)|.
\end{align*}
Gronwall's inequality yields $|\partial_xP(s,t,x,p)|\le C$.
\end{proof}

\section{Estimates of the Derivatives of the Fields}

Before we are going to estimate the derivatives of the fields, we have to recall the corresponding representation formulas for the fields.

\begin{lemma}\label{representationfirstt}
Let $(f,\phi)$ be a solution of (\ref{waveequation}), (\ref{vlasovequation}) and (\ref{initialconditions}) on some time interval $[0,T[$, $T>0$, with initial conditions $(\fin,\phinin,\phiein)\in\mathcal{I}$. Then we have
\begin{displaymath}
\partial_t\phi=(\partial_t\phi)_D+(\partial_t\phi)_T+(\partial_t\phi)_S,
\end{displaymath}
with
\begin{alignat*}{2}
(\partial_t\phi)_D&=&&-\frac{1}{t}\negsp\int_{|x-y|=t}\negsp\int\frac{\fin(y,p)}{1+\omega\SP\hat{p}}\frac{dp}{\sqrt{1+p^2}}\,dS_y\\
&&&+\frac{1}{4\pi t^2}\negsp\int_{|x-y|=t}\negsp\left(\phiein(y)-\partial_x\phiein(y)\SP(x-y)\right)\,dS_y\\
&&&-\frac{1}{4\pi t^3}\negsp\int_{|x-y|=t}\negsp\left(2\partial_x\phinin(y)\SP(x-y)-(\partial_x^2\phinin(y)(x-y))\SP(x-y)\right)\,dS_y,\\
(\partial_t\phi)_T&=&&-\negsp\int_{|x-y|\le t}\negsp\int a^{\phi_t}(\omega,p)f\ret\,dp\frac{dy}{|x-y|^2}\\
&:=&&-\negsp\int_{|x-y|\le t}\negsp\int a^{\phi_t}(\omega,p)f(t-|x-y|,y,p)\,dp\,\frac{dy}{|x-y|^2},\\
(\partial_t\phi)_S&=&&-\negsp\int_{|x-y|\le t}\negsp\int b^{\phi_t}(\omega,p)((S\phi)f)\ret\,dp\frac{dy}{|x-y|}\\
&&&-\negsp\int_{|x-y|\le t}\negsp\int c^{\phi_t}(\omega,p)\SP(\partial_x\phi f)\ret\,dp\frac{dy}{|x-y|},
\end{alignat*}
where the kernels satisfy the estimates
\begin{displaymath}
|a^{\phi_t}(\omega,p)|\le c(1+p^2),\quad |b^{\phi_t}(\omega,p)|\le c\sqrt{1+p^2},\quad |c^{\phi_t}(\omega,p)|\le c.
\end{displaymath}
Here $\omega:=-\frac{x-y}{|x-y|}$ and $\cdot\ret$ means that we evaluate the integrands at retarded time $t-|x-y|$ and at position $y$. For $\partial_{x_i}\phi$ a similar formula holds.
\end{lemma}
\begin{proof}
Up to the term $(\partial_t\phi)_D$ the proof can be found in \cite{calogerorein03}, Proposition~1. The second and third term in $(\partial_t\phi)_D$ correspond to the derivative of the solution $\phi_0$ of the homogeneous wave equation with initial data $\phinin,\phiein$ given by
\begin{displaymath}
\phi_0(t,x)=\frac{1}{4\pi t^2}\negsp\int_{|x-y|=t}\negsp\left(\phinin(y)-\partial_x\phinin(y)\SP(x-y)\right)\,dS_y+\frac{1}{4\pi t}\negsp\int_{|x-y|=t}\negsp\phiein(y)\,dS_y.
\end{displaymath}
The first term in $(\partial_t\phi)_D$ we have just copied from \cite{calogerorein03}, Proposition 1.
\end{proof}

\begin{lemma}\label{representationsecondt}
Let $(f,\phi)$ be a solution of (\ref{waveequation}), (\ref{vlasovequation}) and (\ref{initialconditions}) on some time interval $[0,T[$, $T>0$, with initial conditions $(\fin,\phinin,\phiein)\in\mathcal{I}$. Then we have
\begin{displaymath}
\partial_t^2\phi=(\partial_t^2\phi)_{DD}+(\partial_t^2\phi)_R+(\partial_t^2\phi)_{TT}+(\partial_t^2\phi)_{TS}+(\partial_t^2\phi)_{SS},
\end{displaymath}
with
\begin{alignat*}{2}
(\partial_t^2\phi)_{DD}&=&&\text{{ }}\partial_t(\partial_t\phi)_D-\frac{1}{t^2}\negsp\int_{|x-y|=t}\negsp\int a^{\phi_t}(\omega,p)\fin(y,p)\,dp\,dS_y\\
&&&-\frac{1}{t}\negsp\int_{|x-y|=t}\negsp\int b^{\phi_t}(\omega,p)\fin(y,p)(\phiein(y)+\hat{p}\SP\partial_x\phinin(y))\,dp\,dS_y\\
&&&-\frac{1}{t}\negsp\int_{|x-y|=t}\negsp\int c^{\phi_t}(\omega,p)\SP\partial_x\phinin(y)\fin(y,p)\,dp\,dS_y\\
&&&+\frac{1}{t^2}\int\negsp\int_{|x-y|=t}\negsp\frac{a^{\phi_t}(\omega,p)\omega\SP\hat{p}}{1+\omega\SP\hat{p}}\fin(y,p)\,dS_y\,dp\\
&&&+\frac{1}{t}\int\negsp\int_{|x-y|=t}\negsp\frac{b^{\phi_t}(\omega,p)\omega\SP\hat{p}}{1+\omega\SP\hat{p}}(\phiein(y)+\hat{p}\SP\partial_x\phinin(y))\fin(y,p)\,dS_y\,dp\\
&&&+\frac{1}{t}\int\negsp\int_{|x-y|=t}\negsp\frac{\omega\SP\hat{p}}{1+\omega\SP\hat{p}}c^{\phi_t}(\omega,p)\SP\partial_x\phinin(y)\fin(y,p)\,dS_y\,dp,\\
(\partial_t^2\phi)_{R}&=&&-\int\int_{|\xi|=1}\frac{a^{\phi_t}(\xi,p)\xi\SP\hat{p}}{1+\xi\SP\hat{p}}\,dS_\xi f(t,x,p)\,dp,\\
(\partial_t^2\phi)_{TT}&=&&\negsp\oint_{|x-y|\le t}\negsp\int a^{\phi_{tt}}(\omega,p)f\ret\,dp\,\frac{dy}{|x-y|^3},\\
(\partial_t^2\phi)_{TS}&=&&\negsp\int_{|x-y|\le t}\negsp\int b_1^{\phi_{tt}}(\omega,p)((S\phi)f)\ret\,dp\,\frac{dy}{|x-y|^2}\\
&&&+\negsp\int_{|x-y|\le t}\negsp\int b_2^{\phi_{tt}}(\omega,p)\SP(\partial_x\phi f)\ret\,dp\,\frac{dy}{|x-y|^2},\\
(\partial_t^2\phi)_{SS}&=&&\negsp\int_{|x-y|\le t}\negsp\int c_1^{\phi_{tt}}(\omega,p)(S(fS\phi))\ret\,dp\,\frac{dy}{|x-y|}\\
&&&+\negsp\int_{|x-y|\le t}\negsp\int c_2^{\phi_{tt}}(\omega,p)\SP(S(f\partial_x\phi))\ret\,dp\,\frac{dy}{|x-y|},
\end{alignat*}
where the kernels are smooth and bounded on any set $\partial B_1(0)\times B_r(0)$, $r>0$. The integral with respect to $y$ in $(\partial_t^2\phi)_{TT}$ is understood as a Cauchy principal value. The corresponding kernel $a^{\phi_{tt}}$ satisfies
\begin{displaymath}
\int_{|\xi|=1}a^{\phi_{tt}}(\xi,p)\,dS_\xi=0,\quad p\in\R^3.
\end{displaymath}
The other second order derivatives of $\phi$ fulfil similar representation formulas with kernels having the same properties.
\end{lemma}
\begin{proof}
The derivation of these formulas is standard, cf.~\cite{calogerorein03,glasseystrauss86}. We only elaborate on some points. We differentiate $\partial_t\phi$ with respect to $t$. The first three integrals in $(\partial_t^2\phi)_{DD}$ arise from the differentiation of $(\partial_t\phi)_T$ and $(\partial_t\phi)_S$ with respect to $t$ in the domain of integration $|x-y|\le t$. Differentiation of $(\partial_t\phi)_T$ and $(\partial_t\phi)_S$ "under the integral" and the identity
\begin{displaymath}
\partial_th\ret=\frac{1}{1+\omega\SP\hat{p}}(Sh-\hat{p}\SP\nabla_yh)\ret,
\end{displaymath}
which holds for every $h\in C^1(\R^7)$, gives another three terms which we treat by integration by parts. For details we refer to \cite{calogerorein03}. The last three integrals in $(\partial_t^2\phi)_{DD}$ and $(\partial_t^2\phi)_R$ emerge from this process; $(\partial_t^2\phi)_R$ was forgotten in \cite{calogerorein03}, Proposition 3. The properties of the kernels follow after a straightforward calculation, cf.~\cite{calogerorein03}, Proposition 3.
\end{proof}

Now we are able to estimate the derivatives of the fields.

\begin{proposition}\label{estimatesfirstderivatives}
For all $C_1,C_2>0$ there exists a $C^\ast>0$ such that the following holds: If $(f,\phi)$ is a solution of (\ref{waveequation}), (\ref{vlasovequation}) on some time interval $[0,T[$, $T>0$, with initial conditions $(\fin,\phinin,\phiein)\in\mathcal{I}$, which satisfies
\begin{align*}
\text{(i)}&\quad\sup\{|p|\,|\,(x,p)\in\supp f(t)\}\le C_1,\\
\text{(ii)}&\quad\meas\supp f(t,x,\cdot)\le C_2(1+t)^{-3},\quad x\in\R^3, t\in[0,T[,
\end{align*}
then we have for all $t\in[0,T[$ and all $x\in\R^3$ with $|x|\le R+t$
\begin{displaymath}
K(t,x)\le C^\ast(1+R+t+|x|)^{-1}(1+R+t-|x|)^{-1}.
\end{displaymath}
\end{proposition}
\begin{proof}
We use the representation formula from Lemma \ref{representationfirstt} and estimate the appearing terms one by one. The estimates for $\partial_{x_i}\phi$ are completely analogous.\\
{\em Estimate for $(\partial_t\phi)_D$:{ }} We split up the sum in the second and third term in $(\partial_t\phi)_D$ and denote the integrals by $I_1,\dots,I_5$: $(\partial_t\phi)_D=I_1+\dots+I_5$. We have
\begin{displaymath}
|I_2|\le Ct^{-2}\negsp\int_{|x-y|=t}\negsp\chi_{B_R(0)}(y)\,dS_y\le Ct^{-2}\min\{R^2,t^2\}\le C(1+t)^{-2}.
\end{displaymath}
In the same way we obtain
\begin{displaymath}
|I_4|\le C(1+t)^{-2},\quad |I_1|,|I_3|,|I_5|\le C(1+t)^{-1}.
\end{displaymath}
For $x\in\R^3$ with $|x|\le R+t$ we have
\begin{displaymath}
(1+t)^{-2}\le C(1+R+t+|x|)^{-1}(1+R+t-|x|)^{-1}.
\end{displaymath}
Consider an $x\in\R^3$ with $|x|\le t-R$ and $|x-y|=t$. Then, $t=|x-y|\le t-R+|y|$ and thus $|y|\ge R$, which implies $I_1=I_3=I_5=0$ for $|x|<t-R$. Furthermore, for $t-R\le|x|\le t+R$ we have
\begin{displaymath}
(1+R+t+|x|)^{-1}(1+R+t-|x|)^{-1}\ge C(1+t)^{-1},
\end{displaymath}
which implies
\begin{displaymath}
|(\partial_t\phi)_D|\le C(1+R+t+|x|)^{-1}(1+R+t-|x|)^{-1}.
\end{displaymath}
{\em Estimate for $(\partial_t\phi)_T$:{ }}Let us define
\begin{displaymath}
\chi(\tau,\lambda):=\begin{cases}1,&\lambda\le R+\tau\\ 0,&\lambda>R+\tau\end{cases}.
\end{displaymath}
Because of (i), (ii), (\ref{fest}) and $f(t-|x-y|,y,\cdot)=0$ for $|y|>R+t-|x-y|$ (cf.~(\ref{fzero})) we have
\begin{align}
|(\partial_t\phi)_T(t,x)|&=\left|\,\int_{|x-y|\le t}\int_{|p|\le C_1}a^{\phi_t}(\omega,p)f\ret\,dp\,\frac{dy}{|x-y|^2}\right|\nonumber\\
&\le C\negsp\int_{|x-y|\le t}\negsp(1+t-|x-y|)^{-3}\chi(t-|x-y|,|y|)\,\frac{dy}{|x-y|^2}\nonumber\\
&\le C\negsp\int_{|x-y|\le t}\negsp(1+R+t-|x-y|+|y|)^{-3}\chi(t-|x-y|,|y|)\,\frac{dy}{|x-y|^2},\label{rest}
\end{align}
where we have used in the last inequality the relation
\begin{displaymath}
(1+t-|x-y|)^{-3}\le C(1+R+t-|x-y|+|y|)^{-3},
\end{displaymath}
which holds for $|y|\le R+t-|x-y|$. In the proof of Lemma 5.9 in \cite{reindiss} (\ref{rest}) has been estimated. This result yields
\begin{displaymath}
|(\partial_t\phi)_T(t,x)|\le C(1+R+t+|x|)^{-1}(1+R+t-|x|)^{-1}.
\end{displaymath}
{\em Estimate for $(\partial_t\phi)_S$:{ }}Again, using the same arguments as for $(\partial_t\phi)_T$ we get
\begin{alignat*}{2}
|(\partial_t\phi)_S|&\le\text{{ }}&&C\negsp\int_{|x-y|\le t}\int_{|p|\le C_1}|S\phi|f\ret\,dp\,\frac{dy}{|x-y|}\\
&&&+ C\negsp\int_{|x-y|\le t}\int_{|p|\le C_1}|\partial_x\phi|f\ret\,dp\,\frac{dy}{|x-y|}\\
&\le&&C\negsp\int_{|x-y|\le t}\frac{\chi(t-|x-y|,|y|)}{(1+R+t-|x-y|+|y|)^3}K(t-|x-y|,y)\,\frac{dy}{|x-y|}.
\end{alignat*}
Combining the preceding estimates as well as the same estimates that hold for $\partial_{x_i}\phi$, we get
\begin{align*}
K(t,x)&\le C(1+R+t+|x|)^{-1}(1+R+t-|x|)^{-1}\\
&+C\negsp\int_{|x-y|\le t}\negsp\frac{\chi(t-|x-y|,|y|)}{(1+R+t-|x-y|+|y|)^3}K(t-|x-y|,y)\,\frac{dy}{|x-y|},
\end{align*}
and Lemma 11 in \cite{glasseyschaeffer88} proves the claim.
\end{proof}

\begin{proposition}\label{estimatessecondderivatives}
For all $C_1,C_2,C_3>0$ there exists a $C^{\ast\ast}>0$ such that the following holds: If $(f,\phi)$ is a solution of (\ref{waveequation}), (\ref{vlasovequation}) on some time interval $[0,T[$, $T>0$, with initial conditions $(\fin,\phinin,\phiein)\in\mathcal{I}$, which satisfies
\begin{align*}
\text{(i)}&\quad\sup\{|p|\,|\,(x,p)\in\supp f(t)\}\le C_1,\\
\text{(ii)}&\quad\meas\supp f(t,x,\cdot)\le C_2(1+t)^{-3},\quad x\in\R^3, t\in[0,T[,\\
\text{(iii)}&\quad ||\partial_xf(t)||_\infty\le C_3,
\end{align*}
then we have for all $t\in[0,T[$ and all $x\in\R^3$ with $|x|\le R+t$
\begin{displaymath}
L(t,x)\le C^{\ast\ast}(1+R+t+|x|)^{-1}(1+R+t-|x|)^{-7/4}.
\end{displaymath}
\end{proposition}
\begin{proof}
We begin the proof with the estimate of $\partial_t^2\phi$. The estimates for the other second order derivatives follow straightforward from representation formulas similar to those of Lemma \ref{representationsecondt}.\\
{\em Estimate for $(\partial_t^2\phi)_{DD}$:{ }}By a lengthy, but straightforward calculation and the same methods as in the proof of Proposition \ref{estimatesfirstderivatives} one gets for $t\in[0,T[$ and $x\in\R^3$ with $|x|\le R+t$
\begin{displaymath}
|(\partial_t^2\phi)_{DD}|\le C(1+R+t+|x|)^{-1}(1+R+t-|x|)^{-7/4}.
\end{displaymath}
{\em Estimate for $(\partial_t^2\phi)_R$:{ }}Using assumption (i) and (ii) together with (\ref{fest}) we get
\begin{align*}
|(\partial_t^2\phi)_R|&\le C\negsp\int_{|p|\le C_1}\negsp f(t,x,p)\,dp\le C(1+t)^{-3}\\
&\le C(1+R+t+|x|)^{-1}(1+R+t-|x|)^{-7/4},
\end{align*}
for $t\in[0,T[$ and $x\in\R^3$ with $|x|\le R+t$.\\
{\em Estimate for $(\partial_t^2\phi)_{TS}$:{ }}We use Proposition \ref{estimatesfirstderivatives} and the assumptions to obtain
\begin{displaymath}
|(\partial_t^2\phi)_{TS}|\le C\negsp\int_{|x-y|\le t}\negsp\frac{(1+R+t-|x-y|+|y|)^{-1}}{(1+t-|x-y|)^3}\chi(t-|x-y|,|y|)\,\frac{dy}{|x-y|^2}.
\end{displaymath}
This is exactly the same term as in the proof of Lemma 5.10 in \cite{reindiss}. There it has been estimated by
\begin{displaymath}
C(1+R+t+|x|)^{-1}(1+R+t-|x|)^{-7/4}.
\end{displaymath}
{\em Estimate for $(\partial_t^2\phi)_{TT}$:{ }}$(\partial_t^2\phi)_{TT}$ has the same structure as $E_{f,TT}$ in \cite{reindiss}, and hence can be estimated (with the help of (iii)) as
\begin{displaymath}
|(\partial_t^2\phi)_{TT}|\le C(1+R+t+|x|)^{-1}(1+R+t-|x|)^{-7/4},
\end{displaymath}
for $t\in[0,T[$ and $x\in\R^3$ with $|x|\le R+t$.\\
{\em Estimate for $(\partial_t^2\phi)_{SS}$:{ }}Assumption (i) furnishes
\begin{align*}
(\partial_t^2\phi)_{SS}(t,x)&=\int_{|x-y|\le t}\int_{|p|\le C_1}c_1^{\phi_{tt}}(\omega,p)S(fS\phi)\ret\,dp\,\frac{dy}{|x-y|}\\
&+\int_{|x-y|\le t}\int_{|p|\le C_1}c_2^{\phi_{tt}}(\omega,p)\SP S(f\partial_x\phi)\ret\,dp\,\frac{dy}{|x-y|}=:I_1+I_2.
\end{align*}
In the following we treat only $I_1$, $I_2$ being similar. With the Vlasov equation in the form
\begin{displaymath}
Sf=\divergence_p\left(\left[(S\phi)p+(1+p^2)^{-1/2}\partial_x\phi\right]f\right)+fS\phi
\end{displaymath}
and the product rule we obtain
\begin{equation}\label{}
S(fS\phi)=\divergence_p\left(\left[(S\phi)p+(1+p^2)^{-1/2}\partial_x\phi\right]f\right)S\phi+f\left((S\phi)^2+S^2\phi\right),
\end{equation}
and a similar expression for $S(f\partial_x\phi)$. By substitution and integration by parts we obtain
\begin{alignat*}{2}
I_1&=&&-\negsp\int_{|x-y|\le t}\int_{|p|\le C_1}\partial_p(c_1^{\phi_{tt}}(\omega,p)S\phi)\left[(S\phi)p+(1+p^2)^{-1/2}\partial_x\phi\right]f\ret\,dp\,\frac{dy}{|x-y|}\\
&&&+\int_{|x-y|\le t}\int_{|p|\le C_1}c_1^{\phi_{tt}}(\omega,p)f\left((S\phi)^2+S^2\phi\right)\Ret\,dp\,\frac{dy}{|x-y|}=:I_{11}+I_{12}.
\end{alignat*}
With $|\partial_p(c_1^{\phi_{tt}}(\omega,p)S\phi)|\le C\,K(t,x)$ in mind we obtain by the same arguments as above
\begin{displaymath}
|I_{11}|\le C\negsp\int_{|x-y|\le t}\negsp\frac{\chi(t-|x-y|,|y|)(1+t-|x-y|)^{-3}}{(1+R+t-|x-y|+|y|)^2(1+R+t-|x-y|-|y|)^2}\,\frac{dy}{|x-y|}.
\end{displaymath}
Again, in the proof of Lemma 5.10 in \cite{reindiss}, this term has been estimated by
\begin{displaymath}
C(1+R+t+|x|)^{-1}(1+R+t-|x|)^{-7/4}.
\end{displaymath}
Obviously,
\begin{displaymath}
|(S\phi)^2(t,x)|\le C\,K(t,x)^2,\quad |(S^2\phi)(t,x)|\le C\,L(t,x),
\end{displaymath}
and hence
\begin{alignat*}{2}
|I_{12}|&\le\text{{ }}&&C\negsp\int_{|x-y|\le t}\negsp\frac{\chi(t-|x-y|,|y|)(1+t-|x-y|)^{-3}}{(1+R+t-|x-y|+|y|)^2(1+R+t-|x-y|-|y|)^2}\,\frac{dy}{|x-y|}\\
&&&+C\negsp\int_{|x-y|\le t}\negsp\frac{\chi(t-|x-y|,|y|)}{(1+R+t-|x-y|+|y|)^3}L\ret\,\frac{dy}{|x-y|}\\
&\le&&C(1+R+t+|x|)^{-1}(1+R+t-|x|)^{-7/4}\\
&&&+C\negsp\int_{|x-y|\le t}\negsp\frac{\chi(t-|x-y|,|y|)}{(1+R+t-|x-y|+|y|)^3}L\ret\,\frac{dy}{|x-y|}.
\end{alignat*}
Combining the previous estimates and applying \cite{glasseyschaeffer88}, Lemma 11 completes the proof.
\end{proof}

\section{Continuous Dependence on Initial Data and Proof of the Main Theorem}

Besides the continuation criterion, a necessary ingredient in the proof of Theorem \ref{maintheorem} is continuous dependence on the initial data of a solution of (\ref{waveequation}), (\ref{vlasovequation}), which will be proved now.
\begin{proposition}\label{continuousdependence}
For all $\varepsilon>0,T>0$ there exists a $\delta>0$ such that for all $(\fin,\phinin,\phiein)\in\mathcal{I}$ the following holds: If $\Deltain<\delta$, then the solution of (\ref{waveequation}), (\ref{vlasovequation}) and (\ref{initialconditions}) exists on $[0,T]$, and $||K(t)||_\infty+||L(t)||_\infty\le\varepsilon$, $t\in[0,T]$.
\end{proposition}
\begin{proof}
Let $\varepsilon,T>0$ be given. Choose a $\delta>0$ with
\begin{displaymath}
\delta<\min\left\{1,\frac{\varepsilon}{2C_1},\frac{1}{2C_1(1+T)^2},\frac{\varepsilon}{2C_2},\frac{1}{2C_2}\right\},
\end{displaymath}
$C_1,C_2>0$ defined as below. Consider a $(\fin,\phinin,\phiein)\in\mathcal{I}$ with $\Deltain<\delta$. Because of Theorem 1 in \cite{calogerorein03} the solution exists on some maximal time interval $[0,T_{max}[$. Define
\begin{displaymath}
T^\ast:=\sup\{t\in[0,T_{max}[\,|\,||K(s)||_\infty\le (1+s)^{-2},{ }s\in[0,t]\}.
\end{displaymath}
Now we estimate the quantity $\mathcal{P}(t):=\sup\{|p|\,|\,(x,p)\in\supp f(s), 0\le s<t\}$ from Lemma 3 in \cite{calogerorein03} as follows
\begin{displaymath}
\mathcal{P}(t)\le 1+R+\int_0^t(1+s)^{-2}(1+\mathcal{P}(s))\,ds,\quad t\in[0,T^\ast[,
\end{displaymath}
and Gronwall's inequality implies
\begin{equation}\label{pestimate}
\mathcal{P}(T^\ast)\le (1+R)e=C.
\end{equation}
Using (\ref{fest}) and (\ref{pestimate}) we go through the proof of Lemma 5 in \cite{calogerorein03} and obtain the better result
\begin{displaymath}
||K(t)||_\infty\le C(1+t)\left[\Deltain+\int_0^t||K(s)||_\infty\,ds\right],\quad t<T^\ast.
\end{displaymath}
Again, using Gronwall's inequality we get
\begin{displaymath}
||K(t)||_\infty\le C_1\Deltain,\quad t<T^\ast.
\end{displaymath}
From the definitions of $\delta$ and $T^\ast$ we conclude
\begin{displaymath}
||K(t)||_\infty\le 1/2(1+t)^{-2},\quad 0\le t\le\min\{T,T^\ast\},
\end{displaymath}
and thus $T<T^\ast\le T_{max}$, i.~e.~the solution exists at least on $[0,T]$ and $||K(t)||_\infty<\frac{\varepsilon}{2}$ there.\\
We now estimate $||L(t)||_\infty$. From Lemma \ref{representationfirstt} we have
\begin{displaymath}
\partial_t^2\phi=\partial_t(\partial_t\phi)_D+\partial_t(\partial_t\phi)_T+\partial_t(\partial_t\phi)_S,
\end{displaymath}
and as in the proofs of Propositions \ref{estimatesfirstderivatives} and \ref{estimatessecondderivatives} we get $|\partial_t(\partial_t\phi)_D|\le C\Deltain$. $\partial_t(\partial_t\phi)_T$ and $\partial_t(\partial_t\phi)_S$ will be estimated as follows:
\begin{align*}
|\partial_t(\partial_t\phi)_T|&\le C\Deltain+C\negsp\int_{|x-y|\le t}\negsp\sup_{p\in\R^3}|\partial_tf(t-|x-y|,y,p)|\,\frac{dy}{|x-y|^2}\\
&=C\Deltain+C\int_0^t\int_{|\xi|=s}\sup_{p\in\R^3}|\partial_tf(t-s,x+\xi,p)|\,dS_\xi\,\frac{ds}{s^2}\\
&\le C\Deltain+C\int_0^t||\partial_tf(s)||_\infty\,ds
\end{align*}
and $|\partial_t(\partial_t\phi)_S|\le |I_1|+|I_2|$ with
\begin{alignat*}{2}
|I_1|&\le\,\,&&C\Deltain+C\negsp\int_{|x-y|\le t}\negsp\int|\partial_tS\phi|f\ret\,dp\,\frac{dy}{|x-y|}\\
&&&\hspace*{0.9cm}+C\negsp\int_{|x-y|\le t}\negsp\int|S\phi||\partial_tf|\ret\,dp\,\frac{dy}{|x-y|}\\
&\le\,\,&&C\Deltain+C\Deltain\int_0^t||L(s)||_\infty\,ds+C\int_0^t||\partial_tf(s)||_\infty\,ds,
\end{alignat*}
and an analogous expression for $|I_2|$. Similar estimates hold for the other second order derivatives. Altogether we have
\begin{equation}\label{Lest}
||L(t)||_\infty\le C\Deltain+C\int_0^t||\partial_tf(s)||\,ds+C\Deltain\int_0^t||L(s)||_\infty\,ds.
\end{equation}
Now we express $\partial_tf$ by the Vlasov equation, $\partial_tf=-\hat{p}\SP\partial_xf+F\SP\partial_pf+4fS\phi$, substitute $f$ by (\ref{frep}) and estimate:
\begin{displaymath}
|\partial_tf|\le C\Deltain\left[|S\phi|+|\partial_x\phi|+|\partial_xX|+|\partial_xP|+|F|(|\partial_pX|+|\partial_pP|)\right].
\end{displaymath}
If we define
\begin{displaymath}
T^{\ast\ast}:=\sup\{t\in[0,T[\,|\,||L(s)||_\infty\le 1,{ }s\in[0,t]\},
\end{displaymath}
then for $t\in[0,T^{\ast\ast}[$ Lemma 4 in \cite{calogerorein03} implies $|\nabla_{(x,p)}(X,P)|\le C$, hence
\begin{displaymath}
|\partial_tf|\le C\Deltain,\quad t\in[0,T^{\ast\ast}[.
\end{displaymath}
From the last inequality and (\ref{Lest}) we conclude
\begin{displaymath}
||L(t)||_\infty\le C\Deltain+C\Deltain\int_0^t||L(s)||_\infty\,ds\le C_2\Deltain,\quad t\in[0,T^{\ast\ast}[.
\end{displaymath}
By the definition of $\delta$ and $T^{\ast\ast}$ we obtain $T^{\ast\ast}=T$ and the estimate for $||L(t)||_\infty$ as stated in the proposition.
\end{proof}

We are now able to give the proof of Theorem \ref{maintheorem}.
\vspace*{0.2cm}

\noindent{\slshape Proof of Theorem \ref{maintheorem}:{ }{ }{ }}Fix an $0<\eta\le 1$ such that the lemmas of Section \ref{freestreamingsection} hold. Consider a solution of (\ref{waveequation}), (\ref{vlasovequation}) and (\ref{initialconditions}) on some time interval $[0,a]$, $a>0$, which satisfies (FSC) with respect to $\eta$ on $[0,a[$. Because of Lemma \ref{supportestimate} condition (i) in Propositions \ref{estimatesfirstderivatives} and \ref{estimatessecondderivatives} is satisfied on $[0,a[$. Let $(x,p_1),(x,p_2)\in\supp f(t)$, $t\in[0,a[$. From Lemmas \ref{characteristicestimates} and \ref{partialxfestimate} it follows that
\begin{displaymath}
|p_1-p_2|\le\frac{|X(0,t,x,p_1)-X(0,t,x,p_2)|}{Ct}\le Ct^{-1},
\end{displaymath}
as well as $|p_1-p_2|\le 2C_1$. Both together imply
\begin{equation}\label{measest}
\meas\supp f(t,x,\cdot)\le C(1+t)^{-3},
\end{equation}
i.e., condition (ii) in Propositions \ref{estimatesfirstderivatives} and \ref{estimatessecondderivatives} is fulfilled. Lemma \ref{partialxfestimate} entails condition (iii) in Proposition \ref{estimatessecondderivatives} and we get for $t\in[0,a[$ and $x\in\R^3$ with $|x|\le R+t$
\begin{align}
K(t,x)&\le C_1(1+R+t+|x|)^{-1}(1+R+t-|x|)^{-1},\label{est1}\\
L(t,x)&\le C_1(1+R+t+|x|)^{-1}(1+R+t-|x|)^{-7/4},\label{est2}
\end{align}
$C_1:=\max\{C^\ast,C^{\ast\ast}\}$. Because of $1/2<\beta<3/4$ there exists a $\hat{T}>0$ such that we have for $t\ge\hat{T}$ and $x\in\R^3$ with $|x|\le R+t$
\begin{align*}
C_1(1+R+t+|x|)^{-1}(&1+R+t-|x|)^{-1}\\
&\le \frac{1}{2}\eta(1+R+t+|x|)^{-\beta}(1+R+t-|x|)^{-\beta},\\
C_1(1+R+t+|x|)^{-1}(&1+R+t-|x|)^{-1}\\
&\le \frac{1}{2}\eta(1+R+t+|x|)^{-\beta}(1+R+t-|x|)^{-\beta-1}.
\end{align*}
Now let $a:=\hat{T}$ and $\varepsilon:=\frac{\eta}{2}(1+2R+2\hat{T})^{-2\beta-1}$. Then, by Proposition \ref{continuousdependence} there exists a $\delta>0$ such that for all $(\fin,\phinin,\phiein)\in\mathcal{I}$ with $\Deltain<\delta$ the solution exists on $[0,\hat{T}]$ and $||K(t)||_\infty+||L(t)||_\infty\le\varepsilon$, $t\in[0,\hat{T}]$. By our assumptions on $\varepsilon$ (FSC) with respect to $\eta$ is satisfied on $[0,\hat{T}]$. Define
\begin{displaymath}
T^\ast:=\sup\{t\in[\hat{T},T_{max}[\,|\,\text{On{ }}[0,t]\text{{ }(FSC) with resp.~to{ }}\eta\text{{ }is satisfied}\}.
\end{displaymath}
For $t\in[\hat{T},T^\ast[$ and $x\in\R^3$ with $|x|\le R+t$ we get
\begin{align*}
K(t,x)&\le 1/2\eta(1+R+t+|x|)^{-\beta}(1+R+t-|x|)^{-\beta},\\
L(t,x)&\le 1/2\eta(1+R+t+|x|)^{-\beta}(1+R+t-|x|)^{-\beta-1}.
\end{align*}
From the definition of $T^\ast$ we obtain $T^\ast=T_{max}$. Lemma \ref{supportestimate} and Proposition 2 in \cite{calogerorein2} imply $T_{max}=\infty$. The estimates mentioned in the theorem follow from (\ref{measest}), (\ref{est1}) and (\ref{est2}).
\endproof

\section*{Acknowledgements}
The author thanks his thesis advisor Prof.~Gerhard Rein for a lot of discussions. Without him, this paper would not have been possible.

\end{document}